# Engineering Correlated Transparent Conductors via Metal–Insulator Transition Proximity

**Amit Cohen[1], Nofar Tuval[1], Maria Baskin[1], Tien-Lin Lee[2], Tom Tiwald[3], Scott A. Chambers[4], Sophie D. Beck[5], Lior Kornblum[1]**

[1]Andrew and Erna Viterbi Department of Electrical and Computer Engineering, Technion—Israel Institute of Technology, Haifa 32000-03, Israel

[2]Diamond Light Source Ltd, Harwell Science & Innovation Campus, Didcot OX11 0DE, United Kingdom

[3]J. A. Woollam Co., Inc. 645 M Street, Suite 102, Lincoln, NE68508, USA

[4]Physical and Computational Sciences Directorate, Pacific Northwest National Laboratory, Richland, Washington 99352, USA

[5]Center for Computational Quantum Physics, Flatiron Institute, 162 5th Avenue, New York, New York 10010, USA

*Corresponding Author: amit.cohen@campus.technion.ac.il

## Abstract

Transparent conductors are key functional materials in modern optoelectronic devices. Their performance is governed by an inherent trade-off between optical transparency and electrical conductivity. Correlated metallic oxides offer an attractive workaround to this compromise, where narrow partially-filled bands provide high conductivity, while the absence of visible-range transitions enables optical transparency. Despite this promise, transparency and conductivity remain coupled through carrier density and electronic coherence. Here, we use proximity to a metal-insulator transition (MIT) as a design principle for engineering this balance in correlated-oxide transparent conductors. Epitaxial Ti-substituted $SrVO_3$ films serve as a model platform where the optical window and metallicity are tuned simultaneously. We demonstrate and explain a systematic broadening of the optical window to regimes relevant for silicon photovoltaics. This gain is counterbalanced by carrier depletion and disorder, which drive the loss of coherent transport near the composition-driven MIT. A Haacke figure-of-merit analysis identifies the optimal composition range near the MIT, where the loss of metallic transport is accompanied by persistent near-Fermi-level spectral weight, indicating disorder-assisted localization together with correlation-driven spectral-weight redistribution. These findings establish proximity to an MIT as a general materials-design strategy, using electronic filling as a tuning knob for the transparency-conductivity balance in correlated-oxide transparent conductors.

## Introduction

Transparent conductive oxides (TCOs) are a class of materials that combine two properties: optical transparency and electrical conductivity, making them essential for optoelectronic technologies such as solar cells, light-emitting diodes, and optical sensors.[1] This functionality is fundamentally contradictory, as conventional metals are opaque, while glasses typically exhibit high resistance.

To achieve this desired combination, a common practice is to employ wide-band gap semiconductors.[2,3] These materials are highly transparent in their intrinsic (undoped) state, and upon reasonable doping, can reach adequate conductivity for some optoelectronic applications. However, this approach is inherently limited from a fundamental perspective. Within the Drude model, the plasma frequency ($\omega_p$) is governed by the free-carrier spectral weight, $\omega_p \propto \sqrt{\frac{n}{m^*}}$, where n is the carrier density and $m^*$ is the effective mass. Thus, increasing conductivity by adding carriers tends to shift the plasma edge toward the visible or near-infrared (NIR) range, narrowing the optical transmission window and imposing an intrinsic tradeoff between conductivity and transparency. In addition, high dopant concentrations introduce various impurities, reducing mobility and thereby limiting conductivity.[4] Beyond these physical limitations, many of the most widely used TCOs, such as indium tin oxide (ITO), rely on scarce elements, which may compromise cost-effectiveness in technologies such as photovoltaics, in addition to raising environmental concerns related to its mining.[3,5] These combined limitations motivate the search for alternative material platforms that can overcome the intrinsic trade-off between transparency and conductivity while remaining scalable and compatible with modern device integration.

One class of materials extensively investigated as potential TCO candidates is correlated metallic oxides.[6–10] These materials often feature a narrow, partially-filled transition-metal d band that gives rise to metallic transport, while the available optical transitions lie mostly outside the visible range. This combination allows correlated metallic oxides to remain conductive while being largely transparent to visible light, making them an attractive material platform for transparent-conductor design.

Among correlated metallic oxides, $SrVO_3$ (SVO) has emerged as a particularly promising candidate for TCO applications.[7,11–13] SVO is a perovskite-structured correlated metal that combines room-temperature conductivity approaching that of conventional metals with notable transparency in the visible and NIR spectral range.[7,14] Moreover, SVO is composed of abundant elements, offering a more sustainable and potentially cost-effective alternative to indium-based materials. Its low work function is expected to further facilitate efficient electron injection into underlying semiconductors, making it particularly attractive for integration in optoelectronic devices.[15,16] These properties position SVO as a promising earth-abundant platform for next-generation transparent conductors.

Despite these advantages, SVO also presents several significant limitations that hinder its direct implementation as a practical TCO. From a physical standpoint, its optical transparency is not sufficiently high across the visible-NIR spectrum: significant absorption arises near the blue region (~2.8 eV) due to O 2p → V 3d interband transitions, while in the NIR, reflection (associated with its relatively high plasma frequency) further narrows the usable optical window.[7,17] These intrinsic features limit its overall transparency and highlight the need for a strategy to widen the optical window of SVO.

Previous studies have explored B-site substitution as a route to tune the optical transition energy and plasma frequency in SVO-based solid solutions,[18] with Ti-substituted SVO ($SrTi_xV_{1-x}O_3$, STVO) emerging as a particularly relevant correlated oxide platform due to its composition-dependent transport behavior and metal-to-insulator transition (MIT).[19–27] Alloying SVO ($3d^1$) with Ti ($3d^0$) provides a useful route for tuning its optical response: reducing the carrier concentration lowers $\omega_p$, thereby extending transparency into the NIR. In addition, the larger band gap of $SrTiO_3$ (STO, ~3.2 eV), the x = 1 end member of STVO, shifts interband absorption to higher energies, improving visible-range transmittance.[28,29] This optical enhancement, however, comes at the cost of decreased electrical conductivity. Accordingly, we seek the optimal compositional regime where Ti substitution maximizes transparency gains while preserving conductivity, before the loss of coherent metallic transport drives the system towards the insulating regime.[24–26,30] To realize this, we employ oxide molecular beam epitaxy (MBE), a scalable and industry-compatible growth technique capable of producing high-quality epitaxial films with atomically sharp interfaces, to evaluate STVO as a practical transparent conductor platform.[12] Importantly, Ti incorporation also moves the system toward STO, a material with mature epitaxial integration routes on silicon, thereby offering a realistic pathway for integrating correlated oxides with semiconductor platforms.[31–33] In this context, MBE-grown STVO represents a key step toward maturing SVO as an earth-abundant TCO.

## Experimental

Epitaxial $SrTi_xV_{1-x}O_3$ x=0, 0.1, 0.2, 0.3, 0.5 samples were grown on (001) $(LaAlO_3)_{0.3}(Sr_2AlTaO_6)_{0.7}$ (LSAT) 5×5 $mm^2$ double-side-polished substrates (CrysTec) with a Veeco GenXplor oxide MBE system. The growth was done using a shuttered approach based on our previous SVO growth protocols,[34,35] with an oxygen background pressure of $\sim 5 \cdot 10^{-7}$ Torr and substrate temperature of $1000°C$. Each growth cycle consisted of a 60 s SrO shutter step followed by an equivalent B-site shutter sequence of VO → VO + TiO → VO, where the durations were adjusted to set the Ti content. For example, the 20% Ti composition was grown using a B-site sequence of VO for 18 s, VO + TiO for 12 s, and VO for 18 s. The films were grown with nominal thicknesses around 25 nm. X-ray diffraction (XRD) was acquired using a Rigaku SmartLab with $2\theta/\omega$ geometry. Atomic force microscopy (AFM) images were acquired with an Asylum Research/Oxford Instruments Cypher ES in tapping mode. Optical transmission measurements were performed with an Agilent Cary 5000 UV-Vis-NIR spectrophotometer using a universal measurement accessory (UMA). Spectroscopic ellipsometry measurements were conducted using a Woollam M-2000UI, and analysis was performed using the CompleteEASE software. Double-side-polished STO and LSAT substrates (CrysTec) were measured under the same conditions. Electrical characterization was done using a Quantum Design Physical

Property Measurement System (PPMS) DynaCool system in the Van der Pauw geometry. Hard x-ray photoelectron spectroscopy (HAXPES) measurements were conducted at Beamline I09 at the Diamond Light Source with a Scienta Omicron EW4000 hemispherical analyzer operating at a pass energy of 200 eV and an energy resolution of 250 meV.

## Results

### a. Controlled Growth of Epitaxial STVO Thin Films

We begin by establishing the structural quality of the STVO thin films as a platform for a reliable assessment of composition-dependent optical and transport properties. High-resolution XRD patterns confirm the epitaxial growth of all samples (Fig. 1a), with clear (002) reflections and pronounced thickness fringes indicative of high crystallinity and atomically smooth interfaces. The out-of-plane film reflection shifts to lower 2θ with increasing Ti content, consistent with the larger lattice parameter of STO compared to SVO (Table S1).[24] The corresponding unit-cell volume increases systematically across the series (Fig. 1b), reflecting the composition-dependent lattice expansion of the epitaxial films. AFM images further reveal atomically smooth surfaces with a unit-cell-scale roughness (Fig. 1c).

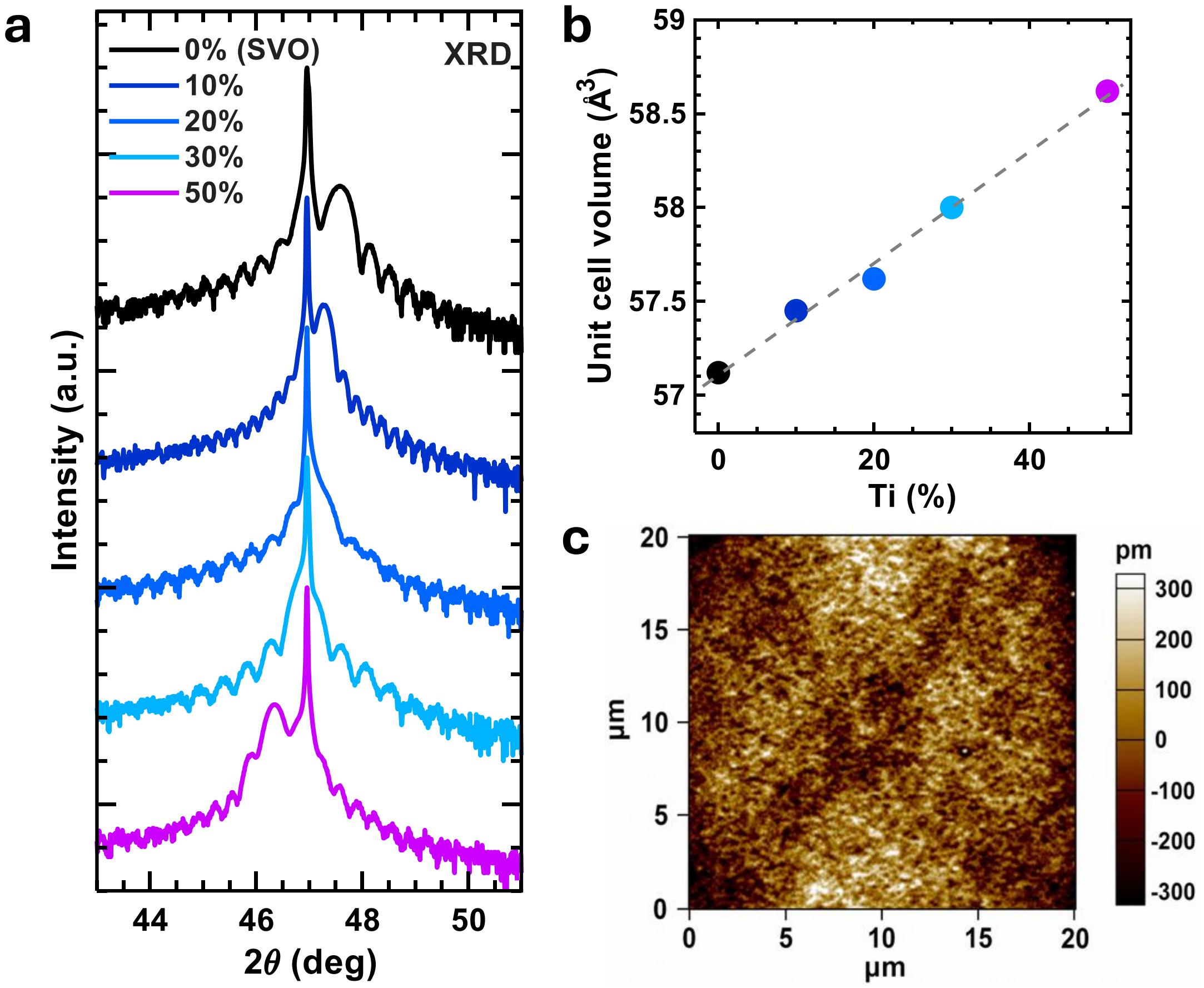

**Figure 1.** Structural characterization of STVO thin films. (a) High-resolution XRD patterns showing the sharp LSAT (002) substrate reflection at ∼ 47° and the composition-dependent STVO (002) film reflections, with thickness fringes observed for all compositions. (b) Unit cell volume as a function of Ti content. (c) Representative AFM topography of the 20% Ti film.

## b. Optimizing STVO as a Transparent Conductor Platform

Having established the structural quality of the STVO series, we next evaluate its transparent-conductor functionality by analyzing the coupled evolution of optical transparency and electrical transport. We begin with the free-standing transmittance of STVO films derived from ellipsometry at a representative thickness of 25 nm, chosen to enable direct comparison across compositions (Fig. 2a). The spectra were obtained from ellipsometry-derived optical constants using an incoherent multiple-reflection model (see Supporting Information).[36] As Ti content increases, systematic spectral shifts and redistribution of spectral weight emerge.

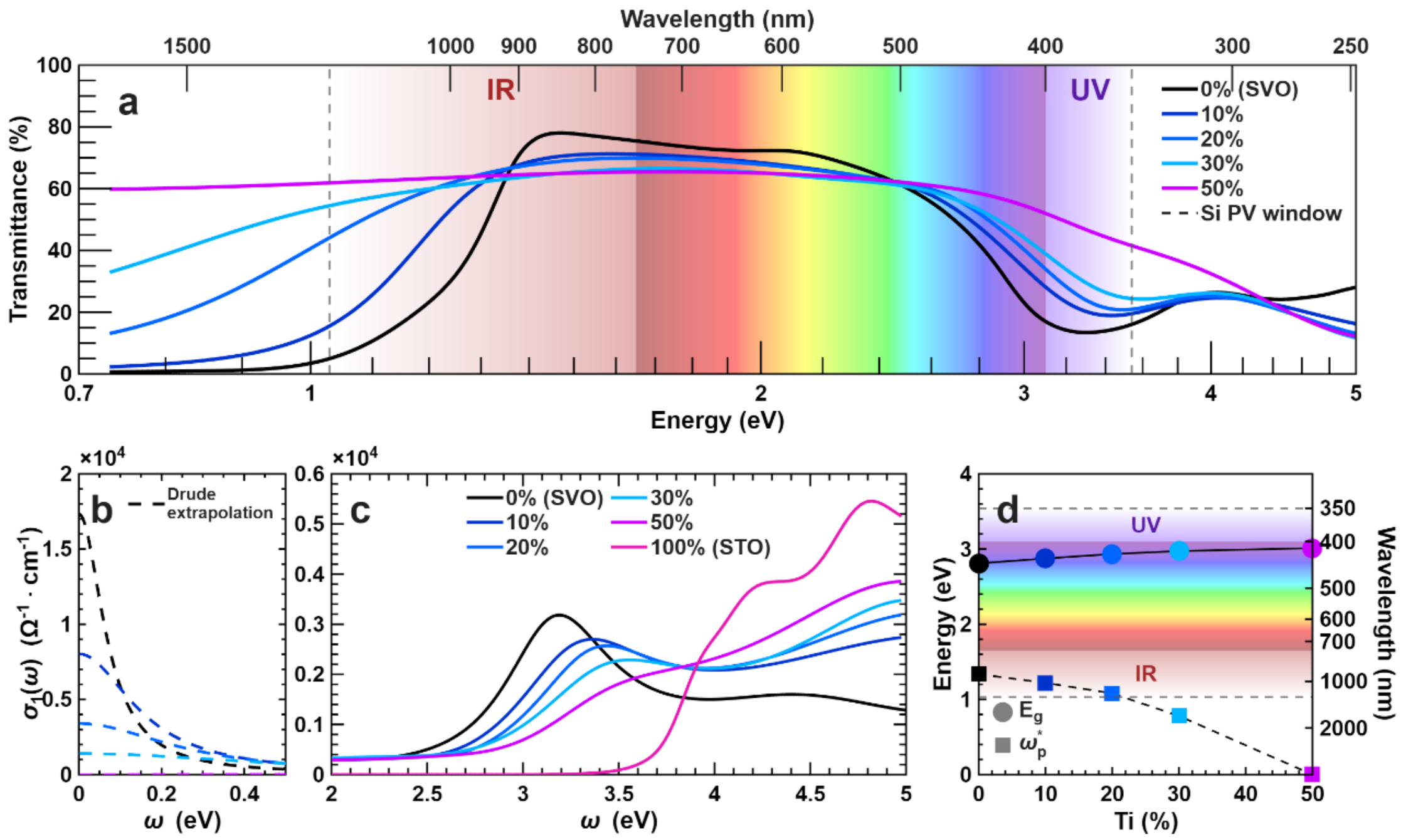

**Figure 2.** Optical properties of STVO thin films as a function of Ti content. (a) Free-standing film transmittance derived for a 25 nm film from ellipsometry-derived optical constants using an incoherent multiple-reflection model. The vertical dashed lines represent the Si solar cell operating window. (b,c) Experimentally derived optical conductivity by recasting the ellipsometry-fitted dielectric functions. The low-energy Drude response in (b) is extrapolated from the measured spectral range shown in (c). (d) Evolution of the O 2p → 3d transition energy (denoted as an effective $E_g$ for convenience, circles), extracted by Tauc analysis, and screened plasma edge ($\omega_p^*$, squares), extracted from the zero crossing of the real part of the dielectric function (see Supporting Information). Horizontal dashed lines mark the Si photovoltaic operating window, as in (a), and connecting lines are guides for the eye.

In the blue region, increasing Ti content shifts the absorption edge to higher energies, from the SVO absorption onset near ~2.8 eV toward the wider-gap STO limit near ~3.2 eV, thereby enhancing transmittance at higher photon energies. The corresponding real part of the optical conductivity, $\sigma_1(\omega)$, obtained from the dielectric function, identifies the origin of this blue-side transparency enhancement as an upward redistribution of interband spectral weight, primarily associated with O 2p → 3d transitions (Fig. 2c), consistent with prior reports.[17,22,37] Density functional theory (DFT) virtual crystal approximation (VCA) optical-conductivity calculations qualitatively capture the upward shift of the O 2p → 3d transition with Ti content (Fig. S2a,b), supporting the assignment of the blue-side optical-window expansion to composition-driven band-structure evolution. Additionally, as the Ti content increases, the ellipsometry-derived optical-conductivity peak around 3eV exhibits substantial broadening and develops a more gradual low-energy edge, corresponding to the transmittance dip at the same energy, consistent with increased B-site disorder.[24] This low-energy edge becomes more abrupt again at the pure STO limit, where all B-site cations are identical, further supporting the role of compositional disorder.

In the red-NIR region, the screened plasma edge, $\omega_p^*$ (namely, the free-carrier plasma response screened by higher-energy electronic polarizability)[38] shifts to lower energies and broadens with increasing Ti content, resulting in a more gradual suppression of the transmittance. This behavior reflects a reduction in the coherent free-carrier spectral weight upon Ti substitution, consistent with the depletion of carrier density caused by replacing nominal $3d^1$ V sites with $3d^0$ Ti sites. The accompanying broadening of the plasma response points to increased carrier scattering upon Ti substitution. Together, with the increase of Ti content, the simultaneous increase in the O 2p → 3d transition energy (labeled $E_g$ in Fig. 2d for convenience) and decrease in $\omega_p^*$ lead to an opening of the optical window, resulting in improved coverage of the silicon solar cell operating range (350-1200 nm, Fig. 2d).

In addition to these edge shifts, the absorption feature around ~1.9 eV is monotonically suppressed with increasing Ti content (Fig. 2a). Its suppression likely reflects the combined effects of reduced V-3d occupancy and correlation-driven redistribution of the remaining 3d states.[17,25,39] Concurrently, an overall reduction in peak transmittance is observed, which we attribute primarily to increased disorder introduced by alloying, likely leading to enhanced scattering and broadening of optical transitions, as discussed below.[25] Direct transmittance measurements reproduce the same spectral evolution at both optical-window edges and the mid-energy absorption feature, validating the ellipsometry-derived trends (Fig. S3).

To assess the conductivity cost associated with optical improvement, we next examine the transport evolution across the STVO series. Pure SVO exhibits clear metallic behavior, with resistivity increasing monotonically with temperature (Fig. 3a), consistent with electron-phonon scattering.[12] The residual resistivity ratio (RRR) of 11.2 further indicates the high quality of the film. With Ti substitution, the room-temperature resistivity increases substantially, and the metallic temperature dependence weakens. These trends are consistent with reduced carrier density as $Ti^{4+}$ ($3d^0$) progressively replaces $V^{4+}$ ($3d^1$), together with enhanced disorder-induced scattering, and correlation-driven loss of electronic coherence, leading to the suppression of coherent metallic transport.[24–26,30] For intermediate compositions, the 10% and 20% Ti samples remain metallic over the measured temperature range, as confirmed by the positive temperature coefficient of resistivity ($d\rho/dT > 0$; Fig. 3b). The 30% Ti sample lies closer to the MIT: it remains metallic at high temperature but undergoes a temperature-driven MIT near ~150 K (Fig. 3c). Consistently, the ratio between the estimated mean free path and the in-plane lattice spacing approaches unity for the 30% Ti sample (Table S3), indicating the breakdown of coherent quasiparticle transport. Beyond this composition-driven crossover, the 50% Ti sample displays insulating behavior over the full measured temperature range, with resistivity decreasing upon warming, marking the loss of metallic conduction.[19,20,24–27] Consistent with this transport evolution, the low-energy optical conductivity is progressively suppressed with Ti substitution (Fig. 2b), indicating a loss of free-carrier (Drude) spectral weight as the metallic state is weakened.

Quantitatively, the Drude-extrapolated zero-frequency resistivity is comparable to the measured 300 K transport resistivity (marked as triangles in Fig. 3a). DFT-VCA optical conductivity reproduces the same qualitative suppression with Ti substitution (Fig. S2a, bottom panel), but substantially underestimates its magnitude. While DFT and experimental values are comparable for SVO, they increasingly diverge with increasing Ti content, indicating that the reduction in carrier density and the average band-structure evolution captured by VCA are insufficient to account for the experimentally observed conductivity suppression (Fig. S2c). This growing discrepancy points to additional composition-dependent mechanisms beyond the VCA description.

Together, the transport and low-energy optical responses establish the conductivity cost of the optical-window expansion, reflecting the combined effects of carrier depletion, electronic correlations, and B-site disorder.[24–26,30] The electronic-structure signatures and roles of these contributions are examined in Sec. 2c.

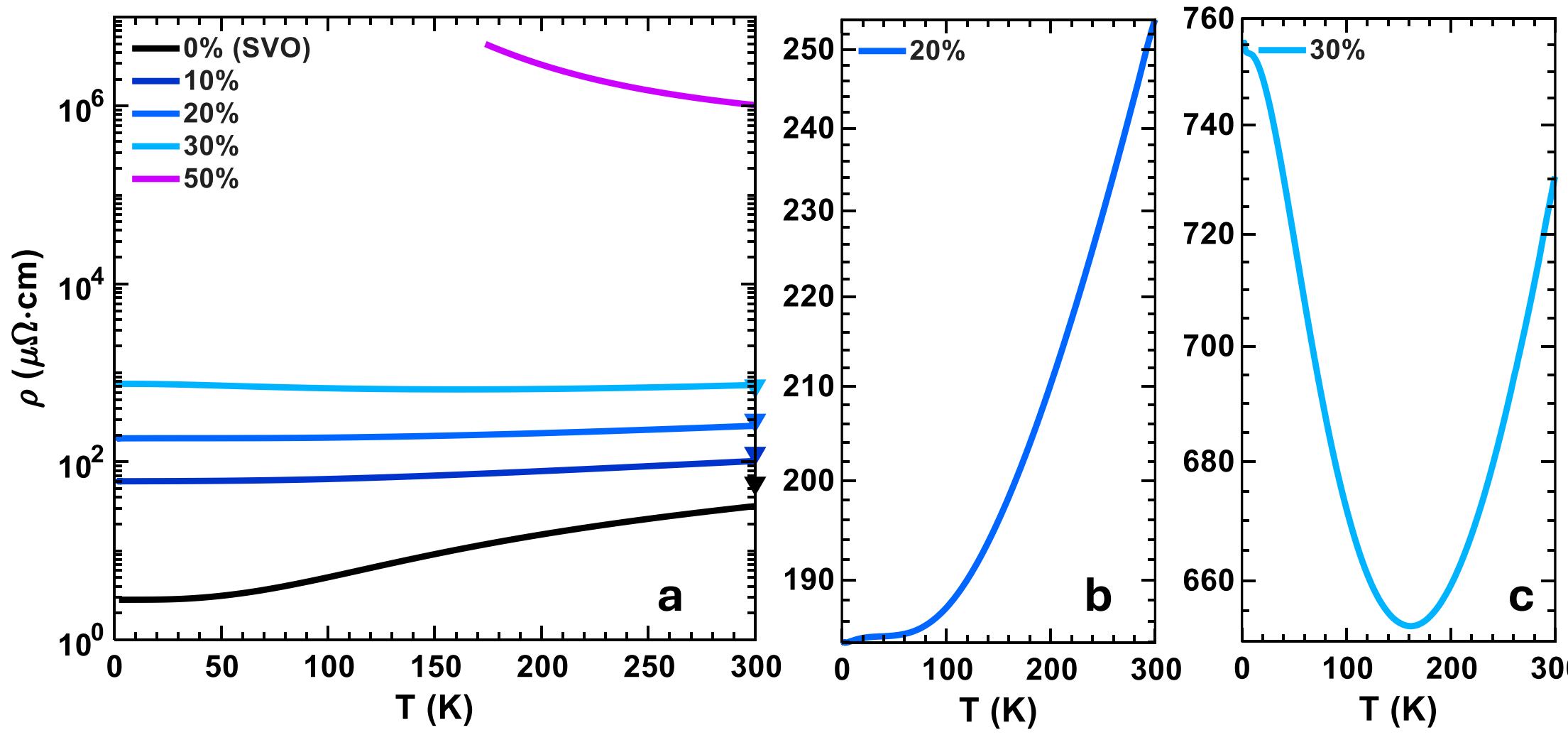

**Figure 3.** Electrical transport properties of STVO thin films. (a) Temperature-dependent resistivity for all compositions, (triangles at 300 K indicate the resistivity value as calculated from the Drude extrapolation of the optical conductivity). Zoomed-in views of the (b) 20% Ti sample showing metallicity across all temperatures and (c) 30% Ti sample where a thermal MIT is observed at ~150 K.

To quantify the combined impact of the opposing optical and transport trends, and to assess the applicability of the STVO series as a TCO, we apply the Haacke figure of merit (FOM),[40] $\phi = T^{10}/R_S$, where T is the transmittance and $R_S$ is the sheet resistance. Because each Ti composition modifies the transparency across different regions of the spectrum (Fig. 2a), the relative ranking of the compositions can depend on the selected wavelength window and weighting function. We therefore employ an application-specific form of this FOM, in which the wavelength-dependent transmittance $T(\lambda)$, is averaged over the relevant wavelength ($\lambda$) range, using a suitable weighting function (W):

$$T_{avg} = \frac{\int_{\lambda_{min}}^{\lambda_{max}} T(\lambda) W \, d\lambda}{\int_{\lambda_{min}}^{\lambda_{max}} W \, d\lambda}$$

Here, $T_{avg}$ is evaluated over the silicon photovoltaic operating window, 350-1200 nm, using the AM1.5g solar spectrum as the weighting function W.[41–44]

We first evaluate the FOM at a representative nominal film thickness of 25 nm to enable direct comparison of the composition-dependent optical and electrical response (Fig. 4a,b). At this thickness, $T_{avg}$ increases from SVO to the 20% Ti sample, indicating improved optical performance upon Ti substitution (Fig. 4a). This improvement reflects the combined benefit of shifting the absorption edge to higher energies and $\omega_p^*$ to lower energies. However, the trend breaks down for the 30% Ti sample, which, despite further edge shifts, exhibits broad disorder-induced optical losses across the spectrum that reduce its overall optical benefit (Fig. 2a). In parallel, $R_S$ increases with Ti content, consistent with the progressive suppression of metallic transport observed in Fig. 3. The resulting FOM reaches its maximum for the 20% Ti sample, identifying the optimal composition in this representative-thickness comparison.

Since both T and $R_S$ depend strongly on film thickness (t), we evaluate the FOM as a function of thickness to identify the optimal operating point of each composition. The corresponding thickness-dependent $R_S$ is estimated using the Mayadas-Shatzkes model (see Supporting Information).[45] The $\phi(t)$ curves exhibit a pronounced maximum with thickness, reflecting the tradeoff between high Rs in the thin-film limit and reduced transmittance at larger thicknesses (Fig. 4d). At low thicknesses, the FOM is primarily limited by sheet resistance, favoring the more conductive SVO and 10% Ti compositions, while at larger thicknesses, optical absorption becomes increasingly limiting, shifting the advantage toward the more transparent 20% and 30% Ti compositions. Near the optimum-thickness regime, which lies in the tens-of-nanometers range for all compositions, the FOM follows the same composition trend observed in the 25 nm comparison: the 10% and 20% Ti samples achieve the highest values, defining the composition range where the transparency–conductivity balance is most effectively optimized. The extracted optimal thickness increases with Ti content (Fig. 4e), reflecting the concurrent reduction in conductivity and increase in transparency: higher Ti compositions can tolerate larger thicknesses due to reduced absorption, but require thicker films to compensate for increased $R_S$. As recently argued, the Haacke FOM does not handle thickness and necessary device performance optimally, but it is used here as a reference performance metric given its prevalence.[46]

Thus, proximity to the MIT provides a practical composition-tuning knob for balancing optical-window expansion against the loss of metallic transport, enabling STVO to be operated near its optimized transparent-conductor regime.

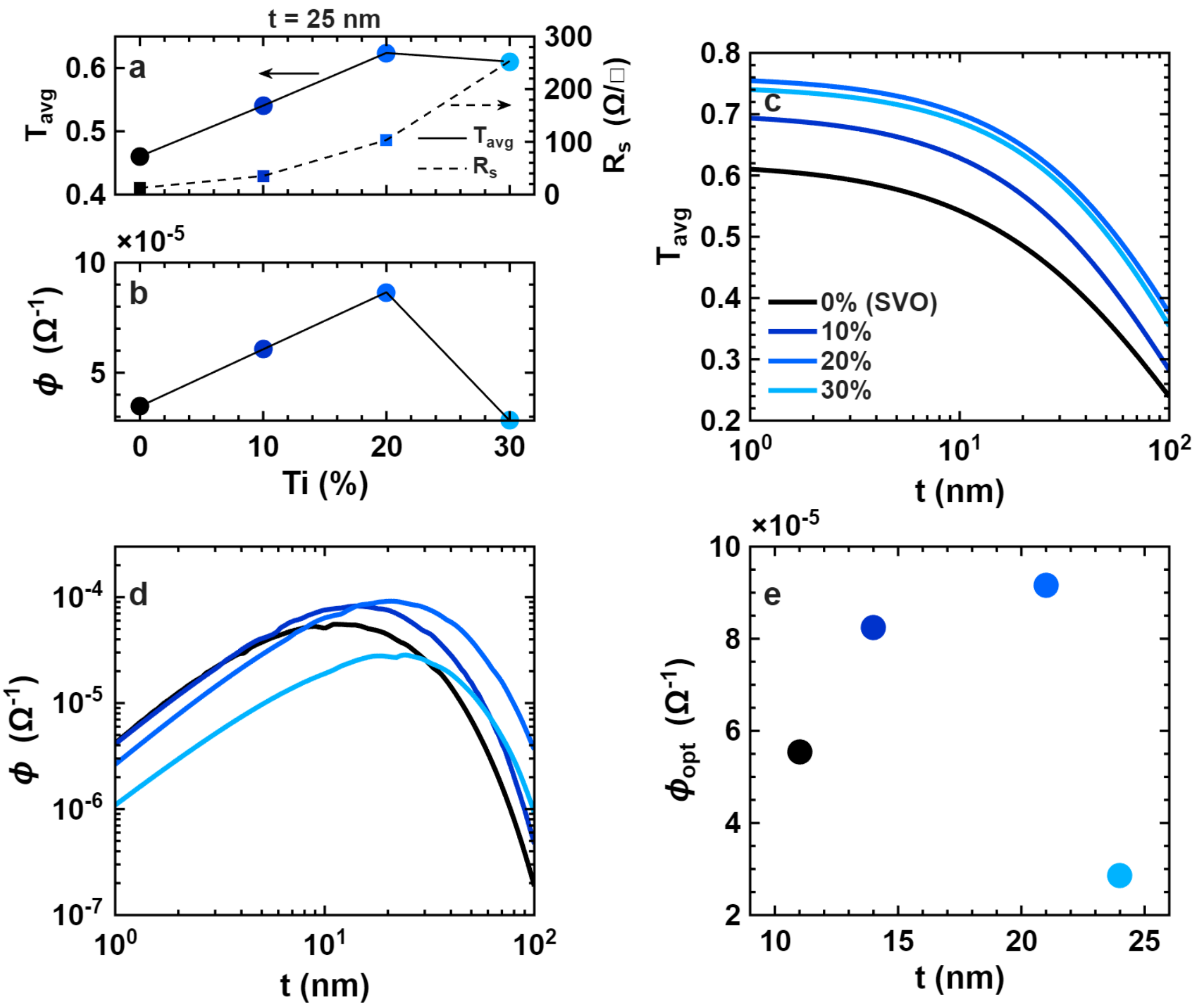

**Figure 4.** Composition and thickness dependence of the FOM for metallic STVO thin films. (a) Weighted average transmittance ($T_{avg}$), and sheet resistance ($R_S$), evaluated at a representative nominal film thickness of 25 nm. $T_{avg}$ is averaged over the silicon photovoltaic spectral window (350–1200 nm) using the AM1.5g solar spectrum as the weighting function. (b) Corresponding FOM, $\phi = T_{avg}^{10}/R_S$, at 25 nm. (c) Thickness-dependent $T_{avg}$. (d) Thickness-dependent FOM. (e) Optimal FOM values.

### c. Electronic Origin of the Transparency-Conductivity Balance

The optical and transport results show that Ti substitution expands the optical window of STVO while progressively suppressing metallic conductivity. We now turn to the electronic origin of this tradeoff. Although Ti substitution connects correlated metallic SVO to conventional band-insulating STO, previous studies have shown that the composition-driven MIT in STVO involves the combined effects of electronic correlations and disorder.[26,39] The Mott-like contribution arises from narrowing of the B-site-derived $t_{2g}$ bandwidth with increasing Ti content, which increases the effective correlation strength and promotes loss of quasiparticle coherence and redistribution of spectral weight toward incoherent states.[25,39] In parallel, random V/Ti occupation of the B-site introduces disorder, providing an Anderson-like localization mechanism for the remaining electronic states.[24,30] The resulting composition-driven MIT has therefore been described as a collaborative Mott-Anderson transition.[26]

To examine how these established mechanisms manifest in our samples, we employ bulk-sensitive HAXPES and DFT-VCA density of states (DOS) calculations. Owing to its large probing depth, HAXPES minimizes surface-related effects[35] and provides access to the occupied electronic structure and core-level final-state response across the STVO series (Fig. 5). DFT-VCA complements this by tracking the composition-dependent evolution of the orbital-resolved density of states (Fig. 6).

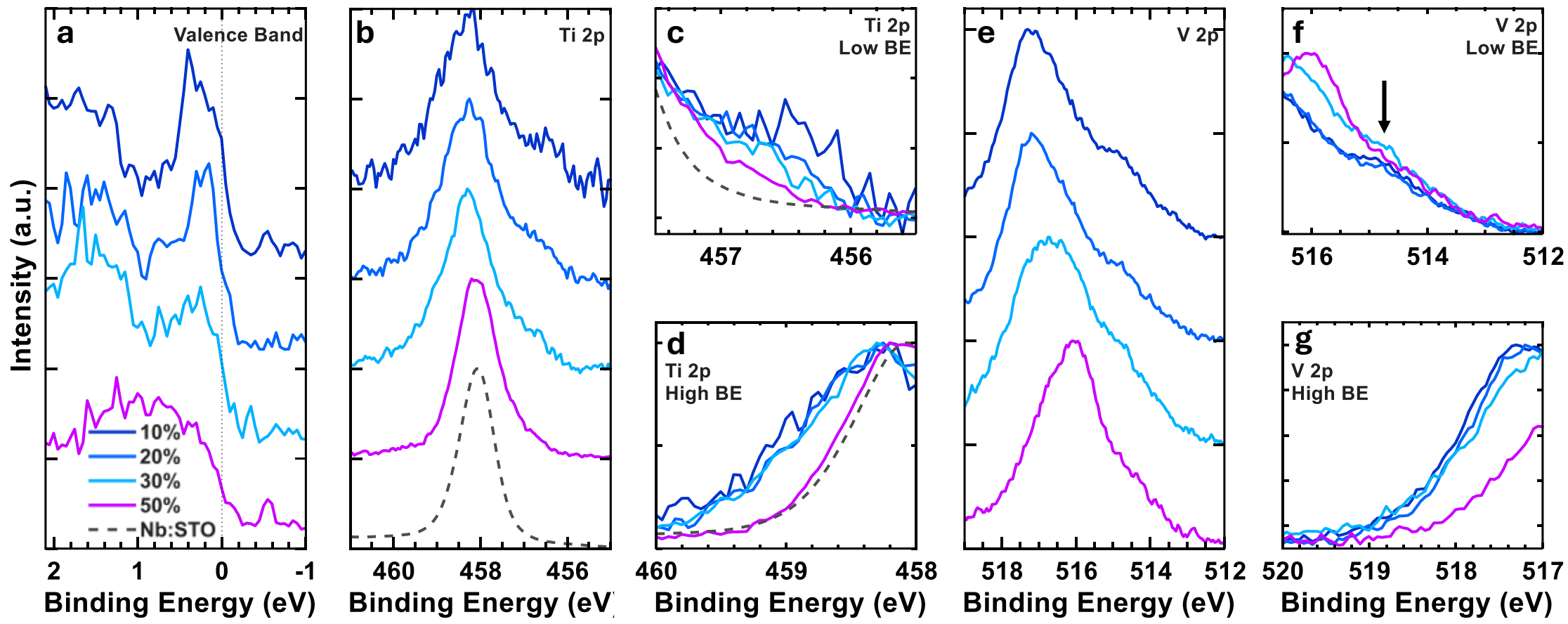

**Figure 5.** HAXPES evolution of the electronic structure across the STVO MIT. (a) Valence band spectra, dashed line denotes $E_F$. (b) Ti 2p core-level spectra. Dashed curve: $SrNb_{0.01}Ti_{0.99}O_3$ (Nb:STO) reference.[47] (c), (d) Expanded views of the Ti 2p low- and high BE features, respectively. (e) V 2p core-level spectra. (f,g) Expanded views of the V 2p low- and high BE features, respectively. The arrow in (f) marks the shoulder at ~514.5 eV, attributed to charge-transfer processes involving the QP and Hubbard-band states.

In the valence-band region, increasing the Ti content transfers spectral weight from $E_F$ to the incoherent feature at ~1 eV (Fig. 5a), consistent with previous ultraviolet photoelectron spectroscopy and DFT reports.[25,26] This spectral-weight redistribution provides a clear correlation-related signature: the coherent V-derived states near $E_F$ are progressively depleted as Ti replaces V, while incoherent spectral weight remains at higher BEs. DFT calculations identify the states involved in this process, showing that metallic transport in STVO is governed by B-site 3d $t_{2g}$-dominated states near $E_F$, which are progressively depleted as the Ti content increases (Fig. 6).[25,39] Importantly, the HAXPES valence-band spectra of the insulating 50% Ti sample retain finite spectral weight near-$E_F$. The coexistence of residual QP-like intensity with insulating $d\rho/dT$ and extremely high room-temperature resistivity (Fig. 3a) is a spectroscopic-transport signature of Anderson-type localization: electronic states remain near $E_F$, but they no longer form extended conducting pathways. Within STVO, this localization occurs alongside correlation-driven spectral-weight redistribution, giving rise to a disorder-assisted Mott-Anderson transition.[20,26,48]

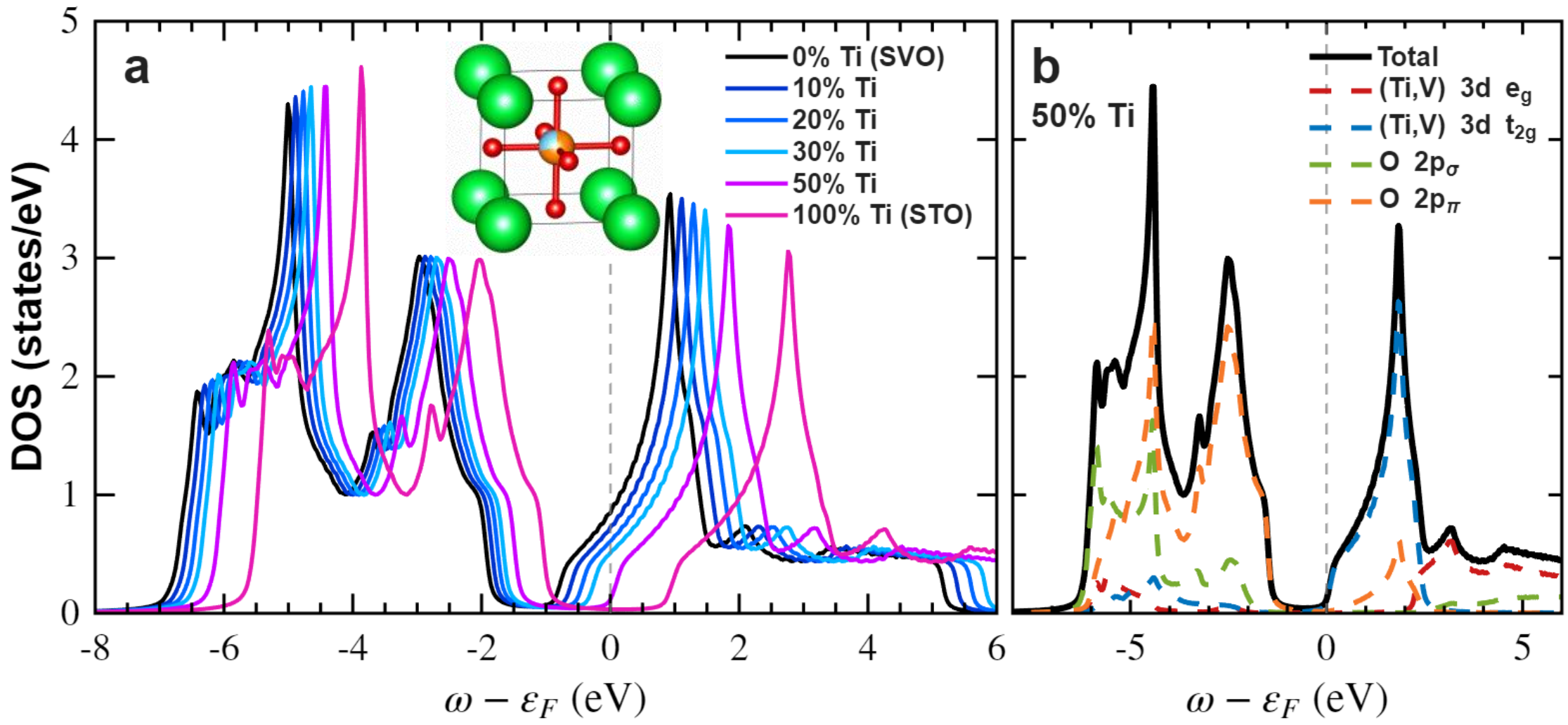

**Figure 6.** Density of states (DOS) calculated by virtual crystal approximation for STVO. (a) DFT-calculated total DOS for the STVO series, showing the evolution from SVO to STO with increasing Ti content. $E_F$ is marked by the dashed vertical line. The inset schematically illustrates the perovskite structure with a 30% Ti / 70% V mixed B-site. (b) Orbital-resolved DOS for the 50% Ti composition.

The Ti 2p spectra provide a local view of how Ti sites couple to the evolving V-derived electronic states (Fig. 5b). With increasing Ti content, the low binding energy (BE) feature near ~456.5 eV is monotonically suppressed (Fig. 5c), indicating a gradual loss of a well-screened, charge-transfer-like Ti final state. This evolution is consistent with Ti substitution reducing the V-derived carrier density and available QP-like spectral weight, thereby weakening local Ti-O-V charge-transfer screening.[26] Although such a low BE feature may formally resemble $Ti^{3+}$-like spectral weight,[49] its systematic suppression with increasing Ti content argues against a simple buildup of chemically reduced Ti sites. We therefore assign it primarily to a final-state screening response that tracks the gradual depletion of V-derived carriers.

In contrast, whereas the low BE feature evolves continuously with Ti content, the high BE feature near ~459 eV is observed exclusively in the metallic samples (10-30%) and is absent in the insulating 50% composition (Fig. 5d). This abrupt metallic/insulating contrast suggests a distinct final-state channel that is lost across the MIT. Its selective presence in the conducting samples implies that this channel is linked to an electronically connected, V-derived metallic network, which is disrupted in the insulating 50% Ti composition. The 50% sample instead exhibits a sharp peak edge closely resembling a $SrNb_{0.01}Ti_{0.99}O_3$ (Nb:STO) reference.[47] There, Ti occupies a $Ti^{4+}$ environment with dilute electron carriers and a valence-band maximum below $E_F$, rather than within a V-derived metallic network. Together, the abrupt disappearance of the high BE feature and the Nb:STO-like line shape support a disorder-assisted loss of electronic connectivity in the V-derived network.

The V 2p spectra provide the complementary view from the conducting sublattice itself. As Ti content increases, the overall V 2p envelope shifts progressively toward lower BEs, reflecting a reorganization of the correlated V final states (Fig. 5e). The shoulder at ~514.5 eV can be associated with charge-transfer processes involving the QP and Hubbard-band states, thus, its weakening reflects the gradual loss of this correlated spectral channel (Fig. 5f).[35,50] This behavior is consistent with the valence-band observation that coherent V-derived spectral weight is progressively depleted with Ti substitution.

The higher-BE region of the V 2p spectrum shows a second, more threshold-like behavior (Fig. 5g). In addition to a smooth weight decrease associated with V dilution and spectral rearrangement, this region exhibits a sharper suppression between 30% and 50% Ti, mirroring the abrupt loss of the Ti 2p high BE feature. Because this abrupt component appears in the V core-level itself, it directly implicates the V-derived conducting network. We therefore assign this sharper change to a disorder-sensitive final-state response that is lost when B-site substitution disrupts the electronic connectivity of the V-O-V network. This provides the core-level counterpart to the transport MIT: spectral weight may remain partially visible in HAXPES, but the extended pathways required for metallic conduction are lost.

Taken together, the valence-band, Ti 2p, V 2p, and DFT results show that the STVO transition is governed by two concurrent processes. First, Ti substitution depletes the V-derived 3d $t_{2g}$ carrier system and redistributes spectral weight away from coherent states near $E_F$, reducing both metallic transport and the free-carrier plasma response. Second, random V/Ti distribution introduces B-site disorder that breaks the connectivity of the V-derived conducting network, localizing the remaining near- $E_F$ states and driving the system into the insulating regime. These effects also provide a natural explanation for the growing divergence between the experimental and DFT-VCA low-energy conductivities discussed above. VCA captures carrier depletion and average band-structure evolution, but these account for only a limited fraction of the observed conductivity suppression, indicating that the dominant additional loss of coherent transport arises from correlation- and disorder-related effects absent from the VCA description. The MIT is therefore best understood as a disorder-assisted Mott-Anderson crossover: correlations contribute to spectral-weight redistribution, while disorder promotes localization and loss of electronic connectivity.

This electronic evolution directly explains the transparency-conductivity tradeoff observed above. The same Ti substitution that expands the optical window by increasing the O 2p → 3d transition energy and lowering the plasma frequency also reduces carrier density, narrows the V-derived bandwidth, and introduces B-site disorder. At intermediate Ti content, the optical window is widened while sufficient coherent transport remains, producing the highest FOM. Closer to the MIT, correlation-driven loss of coherence and disorder-induced localization increasingly suppress conductivity, outweighing further movement toward the wider-gap STO limit.

## Conclusions

In this work, we investigated Ti-substituted SVO as a tunable correlated-oxide platform for transparent-conductor applications. High-quality epitaxial STVO thin films were grown by oxide molecular beam epitaxy, enabling a systematic evaluation of the transparency-conductivity tradeoff near a composition-driven MIT.

Ti substitution expands the SVO optical window through a simultaneous blue shift of the O 2p → 3d absorption edge and red shift of $\omega_p^*$. These changes improve transparency across the spectral range relevant for silicon photovoltaics, but are accompanied by reduced metallic conductivity as carrier depletion, weakened coherent transport, and B-site disorder drive the system toward localization. The combined optical and transport analysis identifies the 20% Ti composition as the optimal balance, yielding the highest Haacke-type FOM for Si solar-cell-relevant operation.

HAXPES and DFT-VCA link this performance limit to a collaborative Mott-Anderson transition, where correlations weaken coherent metallic transport and disorder localizes the remaining carriers across the MIT.

Overall, this study demonstrates that controlled proximity to an MIT can be used to engineer the optical and electrical properties of correlated TCOs. MBE-grown STVO, particularly near $x \approx 0.2$, provides a promising earth-abundant route for expanding the SVO optical window while retaining sufficient conductivity for future oxide/semiconductor optoelectronic integration.

## Acknowledgments

This work was supported by the Israel Science Foundation (ISF Grant 1397/24) and the Israeli Ministry of Innovation, Science & Technology (Grant 1001817852). We acknowledge Diamond Light Source for the beamtime on Beamline I09 under proposal number SI41509-1. The authors thank Lishai Shoham (Technion), for her valuable conceptual input and contribution to the research idea, and to Prof. Marcello J. Rozenberg (CNRS LPS Orsay) for valuable discussions.

## Conflict of Interest

The authors declare no conflict of interest.

## Data Availability Statement

The data that support the findings of this study are available from the corresponding author upon reasonable request.

# Supplementary Section

# Engineering Correlated Transparent Conductors via Metal–Insulator Transition Proximity

**Amit Cohen[1], Nofar Tuval[1], Maria Baskin[1], Tien-Lin Lee[2], Tom Tiwald[3], Scott A. Chambers[4], Sophie D. Beck[5], Lior Kornblum[1]**

[1]Andrew and Erna Viterbi Department of Electrical and Computer Engineering, Technion—Israel Institute of Technology, Haifa 32000-03, Israel

[2]Diamond Light Source Ltd, Harwell Science & Innovation Campus, Didcot OX11 0DE, United Kingdom

[3]J. A. Woollam Co., Inc. 645 M Street, Suite 102, Lincoln, NE68508, USA

[4]Physical and Computational Sciences Directorate, Pacific Northwest National Laboratory, Richland, Washington 99352, USA

[5]Center for Computational Quantum Physics, Flatiron Institute, 162 5th Avenue, New York, New York 10010, USA

*Corresponding Author: amit.cohen@campus.technion.ac.il

## Unit Cell Volume Calculation

The out-of-plane ($a_{\perp}$) lattice parameters were extracted from the X-ray diffraction (XRD) peak positions (Fig. 1a). In-plane lattice parameters were assumed to match that of the $(LaAlO_3)_{0.3}(Sr_2AlTaO_6)_{0.7}$ (LSAT) substrate (3.868 Å), consistent with fully strained epitaxial growth across the $SrTi_xV_{1-x}O_3$ (STVO) series. The corresponding pseudocubic unit-cell volumes were calculated (Table S1).

**Table S1.** Structural parameters of STVO thin films.

| Ti (%) | $a_{\perp}$(Å) | Volume (Å$^3$) |
|---|---|---|
| 0 | 3.818 | 57.12 |
| 10 | 3.840 | 57.45 |
| 20 | 3.851 | 57.62 |
| 30 | 3.877 | 58.00 |
| 50 | 3.918 | 58.62 |

## Free-standing-film Transmittance Calculation

To compare the intrinsic optical response of the STVO films independently of substrate effects, we derived the free-standing film transmittance from the ellipsometry-derived dielectric function.[1] Spectroscopic ellipsometry was first used to extract the complex dielectric function, $\varepsilon(E) = \varepsilon_1(E) + i\varepsilon_2(E)$, for each composition, where E is the photon energy and $\varepsilon_1$ and $\varepsilon_2$ are the real and imaginary components, respectively (Fig. S1). The dielectric response was modeled using a combination of Drude, Lorentz, and Tauc-Lorentz oscillators, accounting for the free-carrier contribution and interband optical transitions (Table S2). The film thicknesses used in the optical model were determined independently by X-ray reflectivity (XRR). The carrier concentrations were obtained from Hall measurements, while the carrier mobilities were calculated from the Hall-derived carrier concentrations and room-temperature resistivity measurements. These independently determined parameters were used in the ellipsometry analysis and are summarized in Table S2. The complex refractive index, $N(E) = n(E) + ik(E)$, was then obtained through

$$n = \sqrt{\frac{\sqrt{\varepsilon_1^2 + \varepsilon_2^2} + \varepsilon_1}{2}},$$

$$k = \sqrt{\frac{\sqrt{\varepsilon_1^2 + \varepsilon_2^2} - \varepsilon_1}{2}}.$$

where n is the refractive index and k is the extinction coefficient. The absorption coefficient was calculated as

$$\alpha = \frac{4\pi k}{\lambda},$$

where λ is the wavelength corresponding to the photon energy. Using the measured film thickness t, the transmittance of an isolated film was calculated within an incoherent multiple-reflection model:

$$T = \frac{(1-R)^2 e^{-\alpha t}}{1 - R^2 e^{-2\alpha t}},$$

where R is the normal-incidence Fresnel reflectance at the air/STVO interface,

$$R = \left|\frac{N-1}{N+1}\right|^2.$$

This approach accounts for absorption within the film together with internal reflections at both film surfaces, while removing substrate contributions, enabling direct comparison between different Ti compositions at a fixed thickness (Fig. 2a).

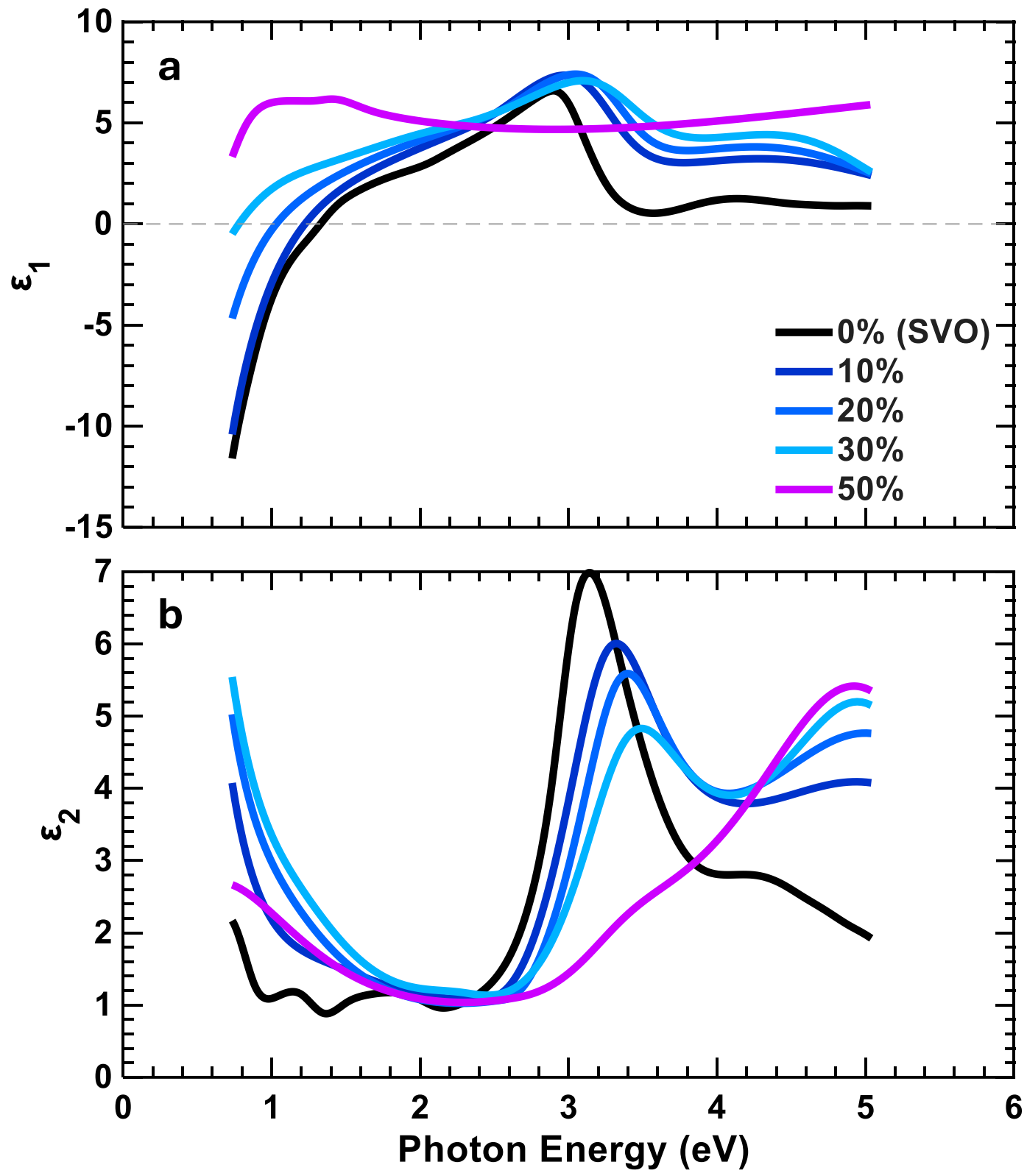

**Figure S1.** (a) Real ($\varepsilon_1$) and (b) imaginary ($\varepsilon_1$) parts of the dielectric function, for the STVO series extracted from spectroscopic ellipsometry.

**Table S2.** Ellipsometry fitting parameters for the STVO series (0–50% Ti), including film thickness, high-frequency dielectric constant $\varepsilon_\infty$, carrier concentration n, carrier mobility μ, effective mass $m^*$, and ultraviolet (UV) pole parameters. The oscillator parameters correspond to the Tauc–Lorentz and Cody–Lorentz contributions used to model the dielectric function. Here, Γ is the oscillator broadening, $E_0$ the resonance energy, $E_g$ the Tauc-Lorentz optical band-gap parameter, and $E_p$ the Cody–Lorentz transition energy parameter. Within the Drude (NMu) parameterization, μ and $m^*$are strongly coupled and should therefore not be interpreted as independently determined quantities. Film thickness, n, and μ were determined independently by XRR, resistivity, and Hall measurements.

| 0% (SVO) | | | | | |
|---|---|---|---|---|---|
| **Thickness (nm)** | **Surface roughness (nm)** | $\varepsilon_\infty$ | $n\ (cm^{-3})$ | $\mu (cm^2V^{-1}s^{-1})$ | $m^*(m_0)$ |
| 23.64 | 1.54 | 2.3 | $2.11 \cdot 10^{22}$ | 5.125 | 3.0 |
| **Oscillator** | **Type** | **Amp.** | **Γ (eV)** | $E_0$ **(eV)** | $E_g$ **(eV)** |
| 1 | Drude | – | – | – | – |
| 2 | Lorentz | 1.1 | 0.7 | 1.8 | – |
| 3 | Lorentz | 1.9 | 1.5 | 4.5 | – |
| 4 | Tauc – Lorentz | 14.4 | 0.7 | 3.2 | 1.4 |
| 5 | Lorentz | 0.7 | 1.0 | 5.5 | – |

| 10% | | | | | | |
|---|---|---|---|---|---|---|
| **Thickness (nm)** | $\varepsilon_\infty$ | $n\ (cm^{-3})$ | $\mu (cm^2V^{-1}s^{-1})$ | $m^*(m_0)$ | **UV Pole Amp.** | **UV Pole Energy (eV)** |
| 41.0 | 1.7 | $2.15 \cdot 10^{22}$ | 2.3 | 3.1 | 121.4 | 11 |
| **Oscillator** | **Type** | **Amp.** | **Γ (eV)** | $E_0$ **(eV)** | $E_g$ **(eV)** | $E_p$ **(eV)** |
| 1 | Drude | – | – | – | – | |
| 2 | Lorentz | 0.6 | 1.4 | 1.6 | – | |
| 3 | Cody – Lorent | 5.5 | 0.9 | 3.3 | 2.5 | 0.5 |
| 4 | Lorentz | 3.7 | 3.0 | 5.3 | – | |

| 20% | | | | | | |
|---|---|---|---|---|---|---|
| **Thickness (nm)** | $\varepsilon_\infty$ | $n\ (cm^{-3})$ | $\mu (cm^2V^{-1}s^{-1})$ | $m^*(m_0)$ | **UV Pole Amp.** | **UV Pole Energy (eV)** |
| 26.9 | 1.8 | $1.5 \cdot 10^{22}$ | 1.4 | 3.2 | 105.4 | 11 |
| **Oscillator** | **Type** | **Amp.** | **Γ (eV)** | $E_0$ **(eV)** | $E_g$ **(eV)** | $E_p$ **(eV)** |
| 1 | Drude | – | – | – | – | |
| 2 | Lorentz | 1.0 | 1.3 | 1.2 | – | |
| 3 | Cody – Lorent | 4.1 | 0.8 | 3.4 | 2.6 | 0.4 |
| 4 | Lorentz | 4.4 | 2.6 | 5.2 | – | |

| 30% | | | | | | |
|---|---|---|---|---|---|---|
| **Thickness (nm)** | $\varepsilon_\infty$ | $n\ (cm^{-3})$ | $\mu (cm^2V^{-1}s^{-1})$ | $m^*(m_0)$ | **UV Pole Amp.** | **UV Pole Energy (eV)** |
| 24.7 | 2.5 | $0.87 \cdot 10^{22}$ | 0.7 | 1.8 | 2.5 | 11 |
| **Oscillator** | **Type** | **Amp.** | **Γ (eV)** | $E_0$ **(eV)** | $E_g$ **(eV)** | $E_p$ **(eV)** |
| 1 | Drude | – | – | – | – | |
| 2 | Tauc – Lorent | 29.6 | 0.8 | 3.3 | 2.3 | |
| 3 | Lorentz | 5.1 | 2.0 | 5.2 | | |

| 50% | | | | | | |
|---|---|---|---|---|---|---|
| **Thickness (nm)** | $\varepsilon_\infty$ | $n\ (cm^{-3})$ | $\mu (cm^2V^{-1}s^{-1})$ | $m^*(m_0)$ | **UV Pole Amp.** | **UV Pole Energy (eV)** |
| 29.8 | – | – | – | – | 285.6 | 10.3 |
| **Oscillator** | **Type** | **Amp.** | **Γ (eV)** | $E_0$ **(eV)** | $E_g$ **(eV)** | $E_p$ **(eV)** |
| 1 | Lorentz | 2.1 | 1.8 | 1.0 | – | |
| 2 | Cody – Lorent | 6.0 | 1.1 | 3.4 | 2.7 | 1.0 |
| 3 | Lorentz | 5.2 | 2.1 | 5.0 | – | |

## Computational Details

### DFT and virtual crystal approximation

Density functional theory calculations were performed with Quantum ESPRESSO[2,3] using the PBE exchange–correlation functional[4] and ultrasoft pseudopotentials from the GBRV library,[5] with plane-wave cutoffs of 60 Ry for the wavefunctions and 720 Ry for the charge density. Two self-consistent calculations were performed for the end members $SrVO_3$ and $SrTiO_3$, using the structures of the Materials Project[6–8] (a = 3.901 Å and 3.912 Å, respectively); note that epitaxial strain is not included in the calculations. The Brillouin zone was sampled with an 11×11×11 Γ-centred mesh using Marzari–Vanderbilt smearing of 0.01 Ry. Fourteen maximally localised Wannier functions[9,10] were created of O 2p and B-site 3d character using Wannier90 with identical projections.[11]

Chemical substitution on the B site was treated within a virtual crystal approximation (VCA) applied at the Wannier level: the real-space Wannier Hamiltonian for composition x was constructed as

$$H_x(R) = (1 - x)H_{SrVO_3}(R) + x\, H_{SrTiO_3}(R).$$

### Density of states and optical conductivity

The density of states, chemical potential, and the optical conductivity were evaluated from $H_x(R)$ by Wannier interpolation using AutoBZ.jl.[12,13] The optical conductivity was obtained from the Kubo formula in the bubble approximation as implemented and described in Ref. [AutoBZ3].[14] The self-energy was taken as a constant imaginary part, $\Sigma = -i\eta$, with $\eta = 50$ meV. The velocity operator was obtained as the Wannier-interpolated band gradient.[15] Calculations used an inverse temperature $\beta = 40\ eV^{-1}$ (T ≈ 290 K) and a uniform 50×50×50 k-mesh, with the same broadening used for the density of states. Vertex corrections are neglected.

### Results

The VCA describes a periodic crystal with a single averaged B site. It therefore contains no B-site disorder: inhomogeneous broadening and Anderson localization, are absent by construction.[16] In particular, the calculated density of states remains metallic for all $x < 1$, including $x = 0.5$, so the calculation provides the disorder-free band-structure reference against which the experimental data is interpreted. Correlation effects like quasiparticle mass enhancement, incoherent spectral features, and any frequency dependence of the scattering are not included. Across the series the O 2p manifold is essentially composition-independent, while the B-site 3d states shift upward relative to it as Ti replaces V, so that the average p–d charge-transfer energy increases with x. The Fermi level moves down within the $t_{2g}$ manifold as the nominal d occupancy decreases from 1 to 1-x. The calculations qualitatively capture the upward shift of the O 2p → 3d peak with increasing Ti content (Fig. S2b, bottom panel), consistent with the experimentally observed blue-side expansion of the optical window (Fig. S2b, top panel). The charge-transfer energy is underestimated as expected by a semilocal functional, as can also be seen in the calculated STO band gap of ~2 eV. We therefore compare trends rather than absolute energies. Second, excitonic effects, which are likely relevant in STO are not captured by the bubble approximation.

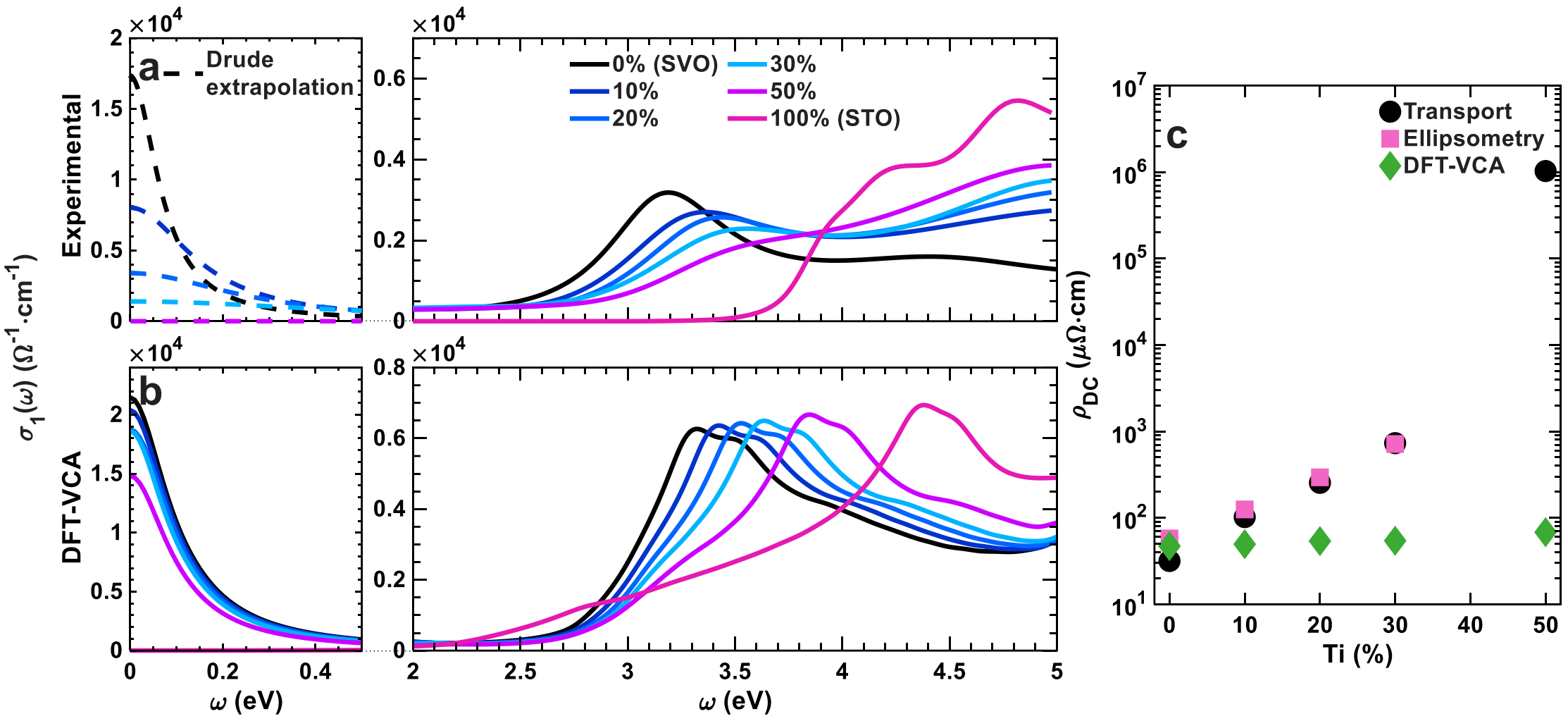

**Figure S2.** Ellipsometry-derived experimental (top) and calculated (bottom) optical conductivity of STVO thin films. (a) Low-energy optical conductivity showing the suppression of the Drude response with increasing Ti content. The top panel dashed lines indicating the Drude extrapolation below the measured spectral range (b) Interband optical conductivity in the O 2p → 3d transition region. (c) Comparison of the DC resistivity obtained from transport measurements at room temperature, Drude extrapolation of the ellipsometry data, and the low-energy DFT-VCA optical conductivity as a function of Ti content.

The calculated zero-frequency conductivity can be compared directly with the measurement. A single broadening is used for the whole series, and it reproduces the SVO conductivity very closely, but increasingly overestimates it as Ti is substituted (Fig. S2c). This is a direct result of disorder, i.e. the scattering that the virtual crystal does not contain introduced by random B-site substitution.

## Experimental Validation of the Ellipsometry-derived Transmittance

To validate the trends extracted from the ellipsometry-based analysis, we compare the calculated free-standing film transmittance with direct transmittance measurements for representative compositions, normalized to the substrate ($T/T_{sub}$, Fig. S3). The measured spectra reproduce the key features observed in the calculated results: a clear blue shift of the high-energy absorption edge with Ti substitution, alongside a shift of the screened plasma edge $\omega_p^*$, to lower energies and a corresponding broadening, resulting in a more gradual decrease in transmittance at low photon energies. In addition, the suppression of the ~1.9 eV feature is consistently observed in both datasets. Quantitatively, the measured transmittance values are in reasonable agreement with the calculated ones. However, the experimental data are presented as normalized transmittance and therefore do not represent an absolute measure of the film-only optical response, failing to incorporate reflection and interface effects.

Taken together, this comparison supports the use of the calculated free-standing film transmittance as a reliable representation of the intrinsic optical response of STVO, while the direct measurements serve as an experimental confirmation of the key spectral evolution with Ti substitution.

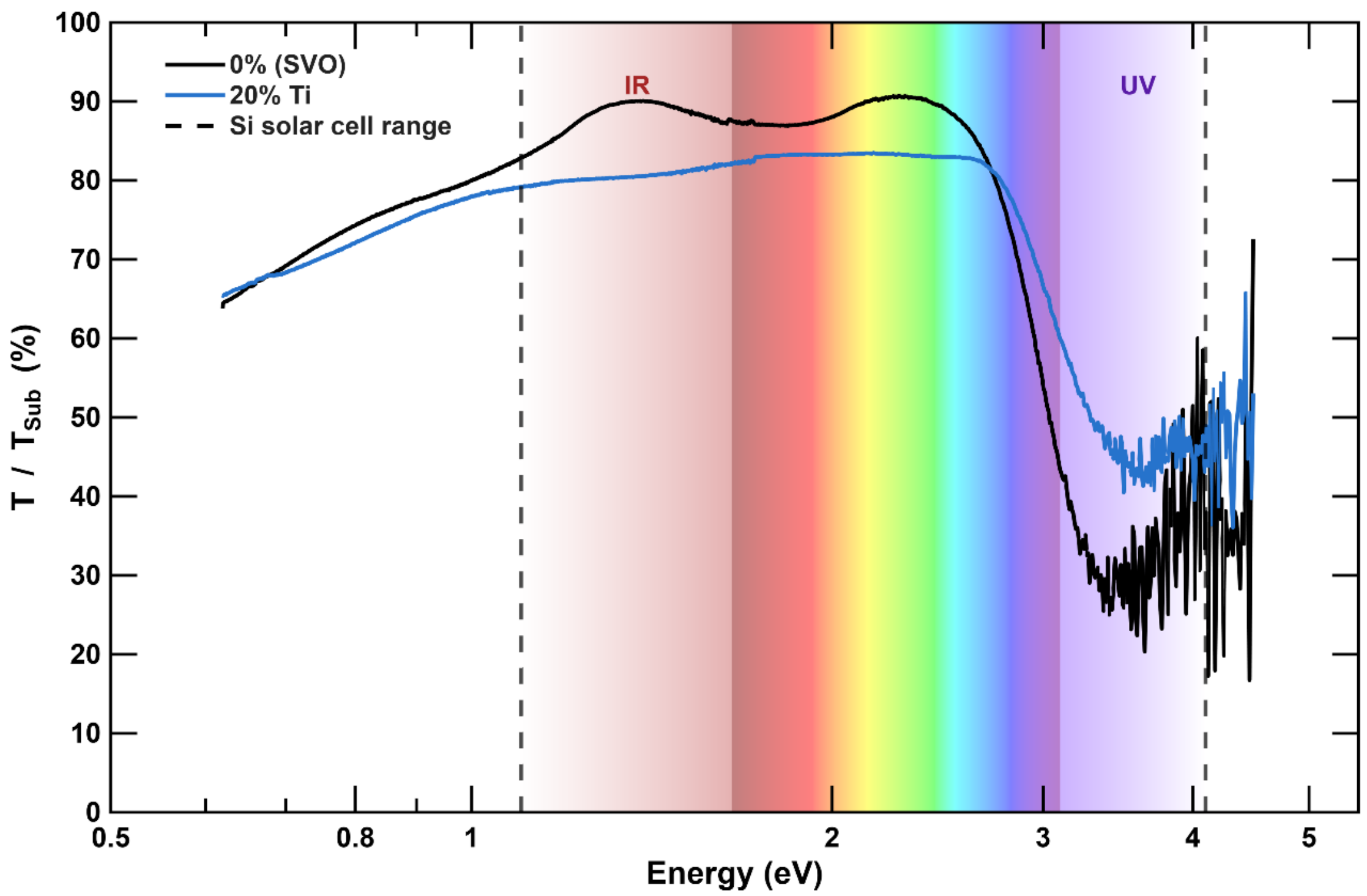

**Figure S3.** Direct optical transmittance of the 0% (SVO) and 20% Ti samples, normalized to the substrate transmission. The dashed lines mark the approximate Si solar-cell spectral range.

## Extraction of $E_g$ and the Screened Plasma Edge From the Dielectric Function

To quantify the evolution of the optical window with Ti substitution, characteristic energies were extracted from the ellipsometry-derived dielectric function (Fig. S1). The O 2p → 3d transition energy is denoted here as $E_g$ for simplicity, although the STVO system remains metallic over part of the composition range.

$E_g$ was extracted using a Tauc analysis of the absorption coefficient obtained from the dielectric function. Both direct and indirect transition models were evaluated (Fig. S4).[17,18] The direct transition energy was used throughout the main text because it exhibits a sharper and more reproducible linear region across the STVO series, enabling more reliable tracking of the composition-dependent spectral evolution. The extracted transition energies increase systematically with Ti content (Fig. S5), consistent with a progressive blue shift of the O 2p → 3d transition.

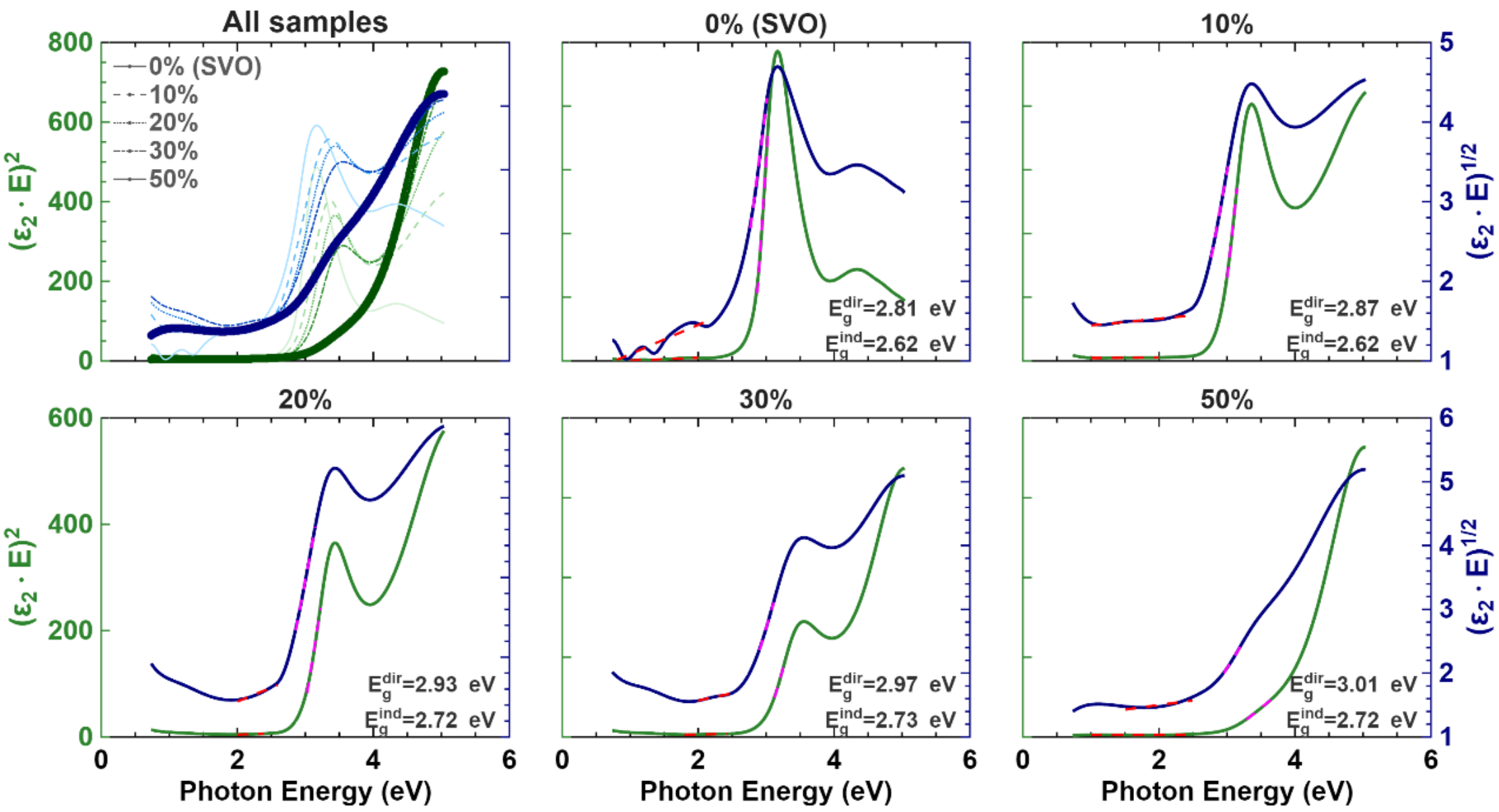

**Figure S4.** Tauc analysis of the ellipsometry-derived absorption coefficient for all STVO compositions, showing the extraction of direct and indirect transition energies.

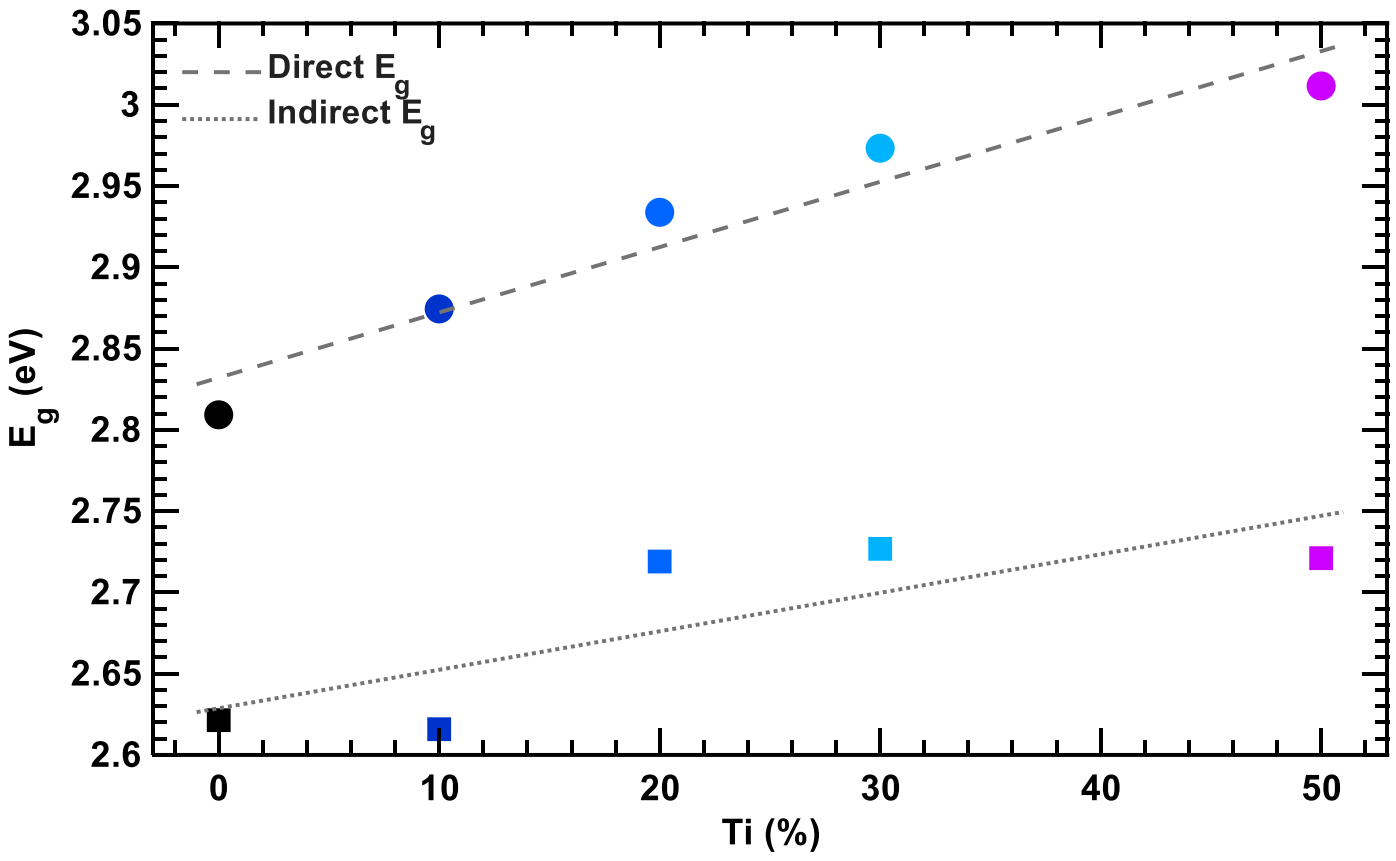

**Figure S5.** Composition dependence of the extracted direct and indirect transition energies. Lines are guides for the eye.

$\omega_p^*$ was determined from the zero crossing of the real part of the dielectric function, $\varepsilon_1(\omega) = 0$ (Fig. S1a).[19] Together, the evolution of $E_g$ and screened plasma-edge captures the simultaneous expansion of the optical window toward both the UV and infrared spectral regions with increasing Ti substitution.

## Thickness-dependent Sheet Resistance Modeling

To estimate the thickness dependence of the sheet resistance of the metallic samples (0-30%), we employed the Mayadas–Shatzkes (MS) formalism, which describes conductivity suppression in thin films due to enhanced electron scattering at surfaces and grain boundaries.[20] The model was used to estimate the thickness-dependent sheet resistance ($R_s(t)$), expected from finite-size scattering effects in the STVO films (Fig. S6).

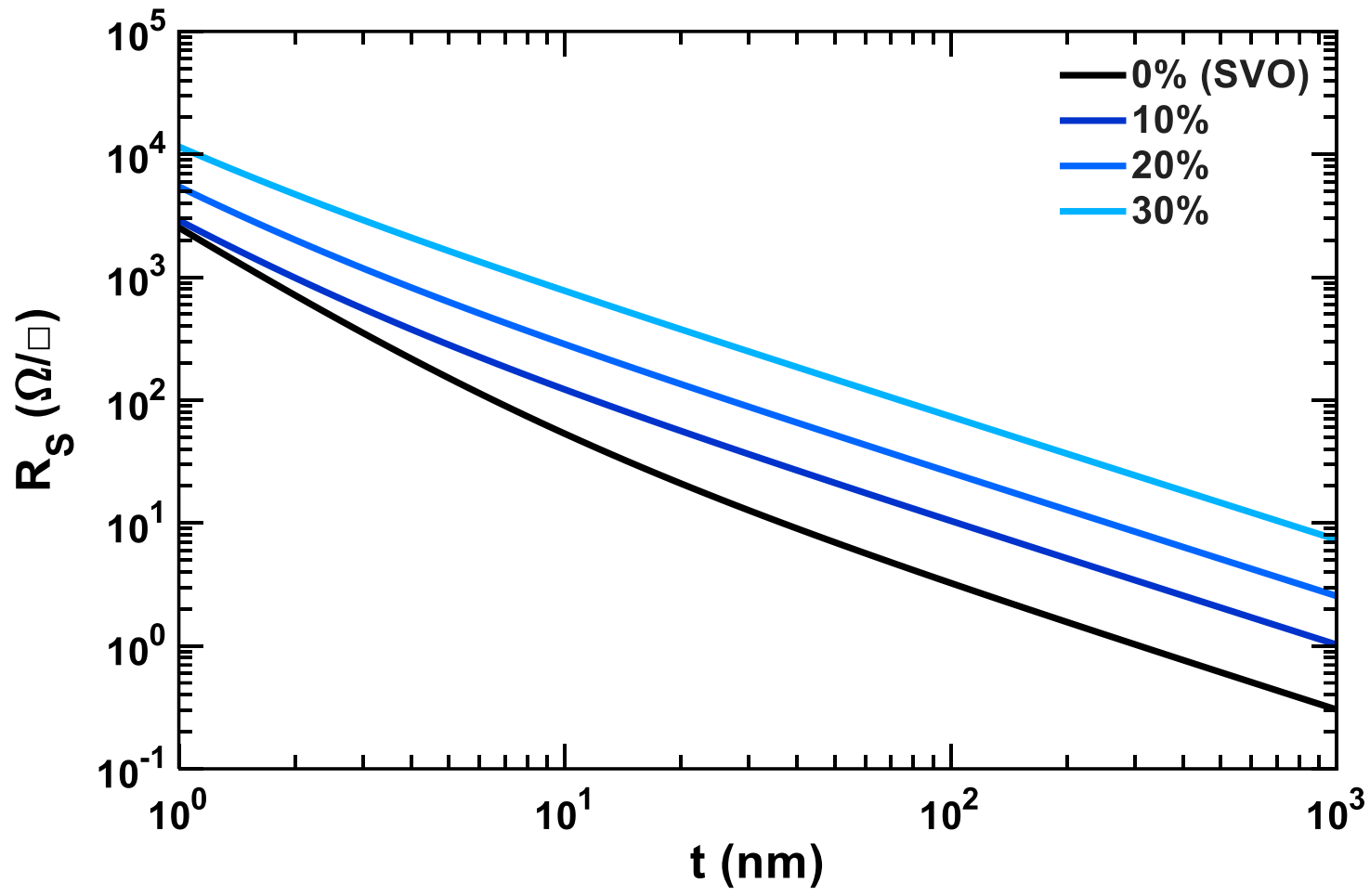

**Figure S6.** Calculated thickness-dependent sheet resistance of STVO films using the Mayadas–Shatzkes model.

The electron mean-free path (Λ, EMFP), was estimated within the Sommerfeld free-electron model using the measured resistivity (ρ), and carrier concentration (n):

$$\Lambda = \frac{\hbar}{e^2\rho}(3\pi^2)^{1/3}n^{-2/3},$$

where $\hbar$ is the reduced Planck constant and e is the elementary charge.[21] The resulting EMFP values decrease strongly with Ti substitution and remain below the film thicknesses considered here (Table S3).

**Table S3.** Carrier concentration n, room-temperature resistivity $\rho_{300K}$, estimated electron mean free path (EMFP) for the STVO series, and $l/a_{||}$ is the ratio between EMFP and the in-plane lattice parameter.

| Ti (%) | $n\ (cm^{-3})$ | $\rho_{300K}$ (μΩ · cm) | EMFP (nm) | $l/a_{\|\|}$ |
|---|---|---|---|---|
| 0 | $2.11 \cdot 10^{22}$ | 30 | 5.55 | 14.35 |
| 10 | $2.15 \cdot 10^{22}$ | 102 | 1.61 | 4.16 |
| 20 | $1.48 \cdot 10^{22}$ | 254 | 0.83 | 2.15 |
| 30 | $8.64 \cdot 10^{21}$ | 730 | 0.41 | 1.06 |

The conductivity correction within the MS formalism is given by

$$\sigma_{MS} = \sigma_0\left[1 - \frac{3}{2}\alpha + 3\alpha^2 - 3\alpha^3 \ln\left(1+\frac{1}{\alpha}\right)\right],$$

With $\alpha = \frac{\Lambda}{D}\frac{R}{1-R}$, where $\sigma_0$ is the bulk conductivity, D is the average grain size, and R is the grain-boundary reflection coefficient. The resulting conductivity was converted into sheet resistance through

$$R_s = \frac{1}{\sigma t},$$

where t is the film thickness. The resulting trends reproduce the enhanced increase in sheet resistance in the ultrathin-film regime expected when the film thickness approaches the EMFP.